\documentclass[prx,twocolumn,amsmath,amssymb,amsthm,superscriptaddress]{revtex4-2}
\usepackage{graphicx,amsmath,relsize,upgreek,color,mathtools,bm,times}
\usepackage[colorlinks=true,linkcolor=blue,citecolor=blue,urlcolor =blue]{hyperref}
\usepackage[hyphenbreaks]{breakurl}
\usepackage{thmtools}
\usepackage{thm-restate}

\newtheorem{lemma}{Lemma}

\newcommand{\ket}[1]{|{#1}\rangle}
\newcommand{\bra}[1]{\langle{#1}|}

\newcommand{\opinner}[3]{\langle #1|#2|#3\rangle}

\newcommand{\rvec}[1]{\pmb{#1}}

\newcommand{\tr}[1]{\mathrm{tr}\!\left\{#1\right\}}

\newcommand{\sinc}{\mathrm{sinc}}

\newcommand{\VAR}[2]{\mathrm{Var}_{#1}\!\left[#2\right]}
\newcommand{\MEAN}[2]{\left<{#1}\right>_{#2}}

\newcommand{\MSE}[2]{\mathcal{D}_{#1}(#2)}

\newcommand{\SPSg}[2]{[\partial_{\mathrm{SPS}}]^{#1}_{#2}}

\newcommand{\SPShod}[3]{[\partial_{\mathrm{SPS}}]^{#1}_{#2}[\partial_{\mathrm{SPS}}]^{#1}_{#3}}
\newcommand{\PSg}[1]{[\partial_{\mathrm{PS}}]_{#1}}

\newcommand{\PShod}[2]{[\partial_{\mathrm{PS}}]_{#1}[\partial_{\mathrm{PS}}]_{#2}}
\newcommand{\FDg}[2]{[\partial_{\mathrm{FD}}]^{#1}_{#2}}

\newcommand{\FDhod}[3]{[\partial_{\mathrm{FD}}]^{#1}_{#2}[\partial_{\mathrm{FD}}]^{#1}_{#3}}
\newcommand{\fQ}{f_{\mathrm{Q}}}

\begin{document}

\title{Robustness of optimized numerical estimation schemes for noisy variational quantum algorithms}

\author{Y.~S.~Teo}
\affiliation{Department of Physics and Astronomy, 
	Seoul National University, 08826 Seoul, South Korea}

\begin{abstract}
  With a finite amount of measurement data acquired in variational quantum algorithms, the statistical benefits of several optimized numerical estimation schemes, including the scaled parameter-shift~(SPS) rule and finite-difference~(FD) method, for estimating gradient and Hessian functions over analytical schemes~[unscaled parameter-shift~(PS) rule] were reported by the present author in~[Y.~S.~Teo, Phys. Rev. A~\textbf{107}, 042421~(2023)]. We continue the saga by exploring the extent to which these numerical schemes remain statistically more accurate for a given number of sampling copies in the presence of noise. For noise-channel error terms that are independent of the circuit parameters, we demonstrate that \emph{without any knowledge} about the noise channel, using the SPS and FD estimators optimized specifically for noiseless circuits can still give lower mean-squared errors than PS estimators for substantially wide sampling-copy number ranges---specifically for SPS, closed-form mean-squared error expressions reveal that these ranges grow exponentially in the qubit number and reciprocally with a decreasing error rate. Simulations also demonstrate similar characteristics for the FD scheme. Lastly, if the error rate is known, we propose a noise-model-agnostic error-mitigation procedure to optimize the SPS estimators under the assumptions of two-design circuits and circuit-parameter-independent noise-channel error terms. We show that these heuristically-optimized SPS estimators can significantly reduce mean-squared-error biases that naive SPS estimators possess even with realistic circuits and noise channels, thereby improving their estimation qualities even further. The heuristically-optimized FD estimators possess as much mean-squared-error biases as the naively-optimized counterparts, and are thus not beneficial with noisy circuits.
\end{abstract}

\maketitle

\section{Introduction}

Quantum computation is a key theoretical milestone of quantum information theory~\cite{Chuang:2000fk} where prospective quantum computers and devices~\cite{Ladd:2010aa,Campbell:2017aa,Lekitsch:2017aa,Arute2019,Zhong:2020quantum,Wu:2021strong} are used to perform tasks with computation power that could in theory surpass classical computers. This prompted the invention of a plethora of quantum-computation and cryptographic algorithms~\cite{Grover:1996fast,Shor:1997polynomial,Raussendorf:2001one-way,Kitaev:2003fault-tolerant,Raussendorf:2007topological,Sehrawat:2011test-state,Montanaro:2016quantum}. In practice, we are still in the era of \emph{noisy intermediate-scale quantum}~(NISQ) devices~\cite{Preskill2018quantumcomputingin} which run algorithms on noisy circuits and a limited number of working qubits~\cite{Bromley:2020applications,Bharti:2022noisy,Finnila:1994quantum,Kadowaki:1998quantum,Aaronson:2011computational,Aaronson:2011linear-optical,Hamilton:2017gaussian,Trabesinger:2012quantum,Georgescu:2014quantum}. These include the class of \emph{variational quantum algorithms}~(VQAs)~\cite{Biamonte:2021universal,Cerezo:2021variational,Cao:2019quantum,Endo:2021hybrid,McArdle:2020quantum} that rely on the interplay between classical and NISQ devices. Examples are quantum eigensolvers designed for quantum-chemistry~\cite{Peruzzo:2014variational,Wecker:2015progress,McClean:2016theory}, combinatorial tasks~\cite{Farhi:2014quantum,Zhou:2020quantum} and quantum machine learning~\cite{Schuld:2015introduction,Schuld:2019quantum,Carleo:2019machine,date2020quantum,Perez-Salinas:2020aa,dutta2021singlequbit,Goto:2021universal,Shin:2023exponential,Shin:2023analyzing}. 

Among the multiple problems faced by NISQ devices, efficient circuit sampling and accurate circuit-function estimation are important goals for achieving practical quantum computation~\cite{Smart:2019quantum-classical,Endo:2021hybrid,vanStraaten:2021measurement}. In recent years, there have been proposals to employ \emph{analytical} estimation schemes, commonly known as the parameter-shift rule~(PS) in the quantum-computing community~\cite{Mitarai:2018quantum,Schuld:2019evaluating,Mari:2021estimating,Wierichs:2022generalparameter}, to exactly estimate gradients and/or Hessians VQAs that rely on, for instance, steepest gradient-descent~\cite{Boyd:2009dv,Fiurasek:2001mq,Rehacek:2007ml,Teo:2011me} and quantum natural gradient-descent methods~\cite{Amari:1998why,Amari:1998natural,Koczor:2019quantum,Stokes2020quantumnatural,Wierichs:2020avoiding} in function optimization. On the other hand, \emph{numerical} estimation schemes are frequently criticized because they are statistically biased and introduce approximation errors. Especially for the finite-difference~(FD) scheme, the general mindset has been that decreasing approximation errors requires a very small step size, and thus a large number of sampling copies to reduce estimation errors. 

Contrary to the above consensus, we note that when circuit functions are to be \emph{sampled}, an FD estimator that gives the minimum \emph{mean-squared error} (MSE, synonymous to ``estimation errors'', or ``sampling errors'' as in~\cite{Teo:2023optimized}) will have an optimized step size that is \emph{not} small---statistical bias is generally necessary to minimize the MSE~\cite{Schonfeld:1971best,Romano:1986counterexamples,Hardy:2002illuminating,Eldar:2004minimum,Shang:2014quantum}. For Pauli-encoded parametrized quantum circuits~(PEPQCs) in which variable parameters are encoded on single-qubit Pauli gates, the present author argued in Ref.~\cite{Teo:2023optimized} that the additional free parameter in a numerical estimator, such as one of those of the FD scheme~(or its generalized versions) or the scaled PS~(SPS) scheme, should be optimized by minimizing the respective MSE averaged over two-design circuits~\cite{harrow_random_2009,Dankert:2009exact,Cleve:2016near-linear}. In situations where barren plateaus exist~\cite{McClean:2018barren,Arrasmith:2021effect,Cerezo:2021cost,Holmes:2022connecting}, that is when the circuit function and its gradient and Hessian magnitudes drop exponentially with the qubit number, these optimized estimators can offer exponentially lower mean-squared errors relative to those from~PS for a \emph{fixed} number of sampling copies.

In this work, we demonstrate that optimized numerical estimators can still be statistically more accurate than analytical ones when quantum circuits are subjected to noise channels, which is part of a crucial research topic that is intimately related to the possibility of a quantum advantage using noisy circuits, especially when the qubit number is large~\cite{Oh:2023classical,Oh:2023tensor,Teo:2023virtual,Hangleiter:2023computational,Aharonov:2023polynomialtime,Deshpande:2022tight,Oh:2021classical,Qi:2020regimes,Noh:2020efficientclassical,GarciaPatron:2019simulatingboson,Oszmaniec:2018classical,Bremner:2017achievingquantum,Kalai:2014gaussian,Aharonov:1996limitations}. After recalling the concepts of gradient and Hessian estimation in Sec.~\ref{sec:numer_analytic} and noisy quantum circuits in Sec.~\ref{sec:noisy_QC}, we first supply closed-form MSE expressions for both the FD and SPS schemes averaged over two-design circuits for a given noise error rate~$\eta$ and qubit number~$n$ in Sec.~\ref{subsec:noisy_MSE}. With these, we show in Sec.~\ref{subsec:naive_numer} that if SPS estimators optimized for noiseless quantum circuits are used to estimate gradient and Hessian components of noisy circuits, then the critical sampling-copy numbers below which these naively-optimized SPS estimators outperform the PS ones all grow as $O(2^n/\eta)$ with increasingly large~$n$ and decreasing~$\eta$. While FD exhibits no closed-form results for these critical numbers, simulations exhibit similar behaviors for both naively-optimized numerical schemes with noisy hardware-efficient circuits.

While the original SPS and FD estimators may be employed when one has absolutely no knowledge about the noise channel, using these naively-optimized numerical schemes that are strictly catered only to noiseless circuits will result in asymptotically wrong estimated gradient and Hessian components. In Sec.~\ref{sec:heuristic}, when \emph{only} the noise-channel error rate is known, we introduce a heuristic error mitigation strategy to reduce the noise biases. Based on the assumptions of unitary two-designs circuits and that the noise-channel error terms do not depend on the circuit parameters, this strategy is independent of the kind of circuit noise channel: it minimizes the MSE \emph{upper bound} over the free parameter for the chosen numerical estimator, which depends only on the error rate and not the noise-channel type. Such a procedure is therefore operational since a very accurate and complete description of the noise channel is unnecessary.

Under this error-mitigation strategy, we find that the heuristically-optimized SPS scheme offers a much more significant reduction in the MSE relative to the naively-optimized SPS and PS schemes even with hardware-efficient quantum circuits and realistic circuit noise. However, the heuristically-optimized FD estimators obtained from this error-mitigation strategy are still as noisily biased as their naively-optimized counterparts due to the way statistical biases enter the approximation errors. This establishes the heuristically-optimized SPS scheme as the preferred choice for noisy gradient and Hessian estimation when the noise-channel error rate is known prior to the estimation.

\section{Numerical and analytical gradient and Hessian estimators for VQAs}
\label{sec:numer_analytic}

A parametrized quantum circuit~(PQC) represented by the unitary operator $U_{\rvec{\theta}}$, where $\rvec{\theta}$ is a collection of real parameters characterizing this circuit of a certain \emph{ansatz}, together with some Hermitian measurement observable~$O$, defines a real circuit function $\fQ(\rvec{\theta})=\opinner{\rvec{0}}{U^\dag_{\rvec{\theta}}OU_{\rvec{\theta}}}{\rvec{0}}$. Here, $\ket{\rvec{0}}\bra{\rvec{0}}=(\ket{0}\bra{0})^{\otimes n}$ is some $n$-qubit pure state initialized to the zero bit-string state of the computational basis. A core purpose of VQAs is to minimize $\fQ(\rvec{\theta})$ over $\rvec{\theta}$. Examples of problems relevant to this task are eigenvalue minimization schemes such as variational quantum eigensolvers~\cite{Peruzzo:2014variational,Wecker:2015progress,McClean:2016theory} and quantum approximate optimization algorithms~\cite{Farhi:2014quantum,Zhou:2020quantum}, where $O$ is a Hamilton operator of either a physical system or combinatorial problem. The PQCs may include classical-data encoding, as in quantum machine learning~\cite{Schuld:2015introduction,Schuld:2019quantum,Carleo:2019machine,date2020quantum,Perez-Salinas:2020aa,dutta2021singlequbit,Goto:2021universal,Shin:2023exponential,Shin:2023analyzing}. Additionally, since the traceless part of $O$ may be written as a sum of traceless Pauli basis operators that are usually each measured independently in an experiment, we shall consider $O$ as a traceless Pauli operator without loss of generality.

In order to present the key results and important messages more easily, we consider \emph{Pauli-encoded parametrized quantum circuits}~(PEPQCs), which are circuits that encode variable parameters on single-qubit Pauli gates defined by the standard Pauli operators $X\equiv\sigma_x$, $Y\equiv\sigma_y$ and~$Z\equiv\sigma_z$. For such PEPQCs and circuits encoded on single-qubit gates of slightly more general Hermitian generators~\cite{Wierichs:2022generalparameter}, one can \emph{exactly} write down the gradient and Hessian components of $\fQ(\rvec{\theta})$. If we consider a rather general and universal circuit \emph{ansatz} consisting of $L$ layers, where each layer comprises single-qubit and two-qubit controlled-NOT~(CNOT) gates, such that $U_{\rvec{\theta}}=W_LW_{L-1}\ldots W_2 W_1$, these are

\begin{align}	
	&\,\PSg{\mu,l}f_{\mathrm{Q}}\equiv\partial_{\mu,l}f_{\mathrm{Q}}=\dfrac{f_{\mathrm{Q}}(\theta_{\mu l}+\pi/2)-f_{\mathrm{Q}}(\theta_{\mu l}-\pi/2)}{2}\,,\nonumber\\
	&\,\PShod{\mu,l}{\mu',l'}\fQ\equiv\partial_{\mu,l}\partial_{\mu',l'}f_{\mathrm{Q}}=\nonumber\\
	=&\,\dfrac{f_{\mathrm{Q}}(\theta_{\mu l}+\frac{\pi}{2},\theta_{\mu' l'}+\frac{\pi}{2})-\fQ(\theta_{\mu l}+\frac{\pi}{2},\theta_{\mu' l'}-\frac{\pi}{2})}{4}\nonumber\\
	&\,-\dfrac{f_{\mathrm{Q}}(\theta_{\mu l}-\frac{\pi}{2},\theta_{\mu' l'}+\frac{\pi}{2})-\fQ(\theta_{\mu l}-\frac{\pi}{2},\theta_{\mu' l'}-\frac{\pi}{2})}{4}\,,
	\label{eq:PS_noiseless}
\end{align}
where the pair $(\mu,l)$ labels the $\mu$th circuit parameter $\theta_{\mu l}$ located in the unitary operator $W_l$. All \emph{other unspecified} parameters of $\fQ$ in the above formulas are otherwise untranslated. The right-hand sides of \eqref{eq:PS_noiseless} constitute the so-called \emph{parameter-shift}~(PS) scheme, which is an \emph{analytical} scheme as it exactly computes the gradient and Hessian components. Since VQAs are iterations of $\fQ$ sampling from a PQC and value updates with a classical computer, these gradient and Hessian components are also estimated from a finite number of sampling copies. We therefore denote the corresponding estimator versions as $\widehat{\PSg{\mu,l}\fQ}$ and $\widehat{\PShod{\mu,l}{\mu',l'}\fQ}$, where each function estimator $\widehat{\fQ}$ is obtained from measuring $N$ copies of the PQC output state in the eigenbasis of a Pauli observable $O$. These estimators therefore possess \emph{finite-copy errors}.

There is another class of \emph{numerical} schemes that approximately defines gradient and Hessian components. One of which is the (centralized) finite-difference~(FD) scheme~(its generalized variants shall not be discussed here):
\begin{align}	
	&\,\FDg{\epsilon}{\mu,l}f_{\mathrm{Q},k}\equiv\sinc(\epsilon/2)\,\partial_{\mu l}f_{\mathrm{Q},k}\nonumber\\
	=&\,\dfrac{f_{\mathrm{Q},k}(\theta_{\mu l}+\epsilon/2)-f_{\mathrm{Q},k}(\theta_{\mu l}-\epsilon/2)}{\epsilon}\,,\nonumber\\
	&\,\FDhod{\epsilon}{\mu,l}{\mu',l'}f_{\mathrm{Q},k}\equiv[\sinc(\epsilon/2)]^2\partial_{\mu,l}\partial_{\mu',l'}f_{\mathrm{Q},k}\nonumber\\
	=&\,\dfrac{f_{\mathrm{Q}}(\theta_{\mu l}+\frac{\epsilon}{2},\theta_{\mu' l'}+\frac{\epsilon}{2})-\fQ(\theta_{\mu l}+\frac{\epsilon}{2},\theta_{\mu' l'}-\frac{\epsilon}{2})}{\epsilon^2}\nonumber\\
	&\,-\dfrac{f_{\mathrm{Q}}(\theta_{\mu l}-\frac{\epsilon}{2},\theta_{\mu' l'}+\frac{\epsilon}{2})-\fQ(\theta_{\mu l}-\frac{\epsilon}{2},\theta_{\mu' l'}-\frac{\epsilon}{2})}{\epsilon^2}\,,
	\label{eq:FD_noiseless}
\end{align}
for $\epsilon>0$. Note that the effective multiplicative factors involving $\sinc(\epsilon/2)=2\sin(\epsilon/2)/\epsilon$ is a special property of PEPQCs, where
\begin{align}
	\fQ(\theta_{\mu,l}+\theta_0)=&\,\fQ(\theta_{\mu,l})+\sin\theta_0\,\partial_{\mu,l}\fQ(\theta_{\mu,l})\nonumber\\
	&\,+(1-\cos\theta_0)(\partial_{\mu,l})^2\fQ(\theta_{\mu,l})\,.
\end{align}
One may also consider a different numerical scheme where a scalar parameter $\lambda$ is multiplied to all true gradient and Hessian components. This nabs us the \emph{scaled parameter-shift}~(SPS) scheme inasmuch as 
\begin{align}	
	\SPSg{\lambda}{\mu,l}f_{\mathrm{Q}}\equiv&\,\lambda\PSg{\mu,l}f_{\mathrm{Q}}\,,\nonumber\\
	\SPShod{\lambda}{\mu,l}{\mu',l'}\fQ\equiv&\,\lambda\PShod{\mu,l}{\mu',l'}\fQ\,.
	\label{eq:SPS_noiseless}
\end{align}
Notice the difference between how the free parameters enter the SPS and FD schemes. 

Unlike the analytical PS scheme, both of these numerical schemes introduce additional \emph{approximation errors} whenever $\epsilon\neq0$ and $\lambda\neq1$. Hence, in VQAs, the estimator counterparts $\widehat{\FDg{\epsilon}{\mu,l}\fQ}$, $\widehat{\FDhod{\epsilon}{\mu,l}{\mu',l'}\fQ}$, $\widehat{\SPSg{\lambda}{\mu,l}\fQ}$ and $\widehat{\SPShod{\lambda}{\mu,l}{\mu',l'}\fQ}$ will also be \emph{statistically biased} (for instance, the data averages $\overline{\widehat{\FDg{\epsilon}{\mu,l}\fQ}}\neq\partial_{\mu,l}\fQ$ and $\overline{\widehat{\SPSg{\lambda}{\mu,l}\fQ}}\neq\partial_{\mu,l}\fQ$) and result in approximation errors. This means that these estimators possess \emph{both} finite-copy and approximation errors.

As a measure for the estimation quality or accuracy, we investigate the mean-squared error~(MSE):
\begin{align}
	\MSE{}{\partial \fQ}=&\,\left<\overline{\left(\widehat{[\partial]_{\mu,l}\fQ}-\partial_{\mu,l}\fQ\right)^2}\right>\,,\nonumber\\
	\MSE{}{\partial\partial \fQ}=&\,\left<\overline{\left(\widehat{[\partial]_{\mu,l}[\partial]_{\mu,l}\fQ}-(\partial_{\mu,l})^2\fQ\right)^2}\right>\,,\nonumber\\
	\MSE{}{\partial\partial' \fQ}=&\,\left<\overline{\left(\widehat{[\partial]_{\mu,l}[\partial]_{\mu',l'}\fQ}-\partial_{\mu,l}\partial_{\mu',l'}\fQ\right)^2}\right>\,.
	\label{eq:MSE_noiseless}
\end{align}
The symbols $\left<\vphantom{W}\hphantom{W}\right>$ and $\overline{\vphantom{W}\hphantom{W}}$ respectively refer to averages over~$\rvec{\theta}$ and measurement data per $\rvec{\theta}$, and $\partial\partial \fQ$ and $\partial\partial' \fQ$ are shorthand for diagonal and off-diagonal Hessian components.

Technically, a numerical estimator would end up with an MSE that is a \emph{sum} of the finite-copy error and approximation error [see Eq.~\eqref{eq:finite_approx_errs}]. On the other hand, an analytical estimator only has the finite-copy error as its MSE. So, why are numerical estimators interesting? Well, in textbook scenarios, they are not when $\fQ$ is computable exactly, in which case $\epsilon=0$ and $\lambda=1$ should be the only sensible options for error-free gradient and Hessian computation. However, they become interesting when $\fQ$ has to be sampled from PQCs. Then, the free parameter~($\epsilon$ or $\lambda$) of a numerical estimator can be chosen as the optimal one that minimizes the MSE. In Ref.~\cite{Teo:2023optimized}, the optimized numerical estimators were shown to give MSEs that drop exponentially in the qubit number~$n$ in the absence of noise. For finite sampling-copy numbers~$N$, the copy-number ranges within which FD estimators outperform PS ones increase exponentially with~$n$. Furthermore, it has also been shown that SPS always outperforms PS for any given~$N$. These results demonstrate that the statistical bias in an estimator, \emph{when optimized properly}, is a key ingredient for minimizing its MSE, an insight well-known in sampling theory~\cite{Schonfeld:1971best,Romano:1986counterexamples,Hardy:2002illuminating,Eldar:2004minimum,Shang:2014quantum}.

\begin{figure}[t]
	\centering
	\includegraphics[width=\columnwidth]{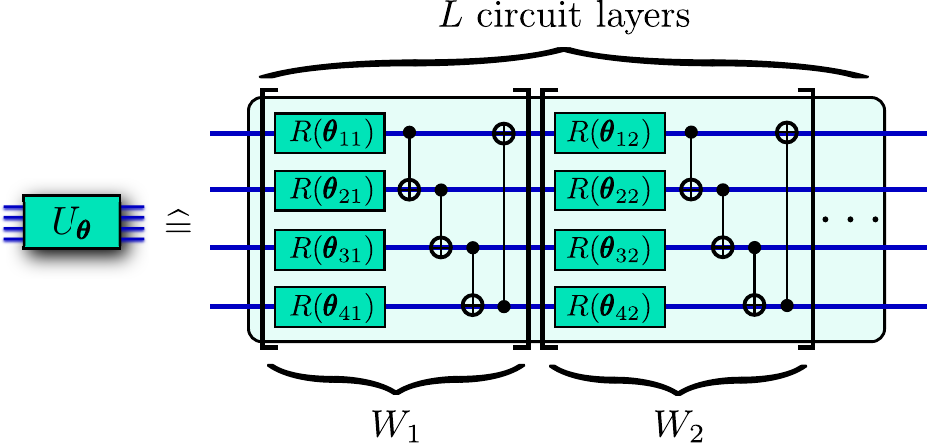}
	\caption{\label{fig:ansatz}Schematic of a four-qubit circuit representation of a $U_{\rvec{\theta}}$ unitary operator. Each $W_l$, or a circuit unitary layer, comprises a chain of parametrized single-qubit rotations ($R$), and CNOT array.} 
\end{figure}

\section{Noisy quantum circuits}
\label{sec:noisy_QC}

Realistic (PE)PQCs are always susceptible to noise in the form of a noise-channel action, namely $\rho_{\rvec{\theta}}=U_{\rvec{\theta}}\ket{\rvec{0}}\bra{\rvec{0}}U_{\rvec{\theta}}^\dag\mapsto\rho'_{\rvec{\theta}}=\mathcal{E}[\rho_{\rvec{\theta}}]$, that is completely-positive and trace-preserving. The resulting noisy mixed state~$\rho'_{\rvec{\theta}}$ can always be written as
\begin{equation}
	\rho'_{\rvec{\theta}}=\rho_{\rvec{\theta},\eta}=(1-\eta)\rho_{\rvec{\theta}}+\eta\rho_\mathrm{err}(\rvec{\theta},\eta)\,,
\end{equation}
with $\eta$ characterizing the error rate, or the strength of the noise-channel map $\mathcal{E}$, and $\rho_\mathrm{err}(\rvec{\theta},\eta)$ is the noise-channel error term that is typically a function of both the noiseless state~$\rho_{\rvec{\theta}}$ and $\eta$, and, thus, also depends on $\rvec{\theta}$.

One example of a realistic noise channel is the successive action of a two-qubit depolarizing channel on an $n$-qubit quantum state $\rho_0$ after every two-qubit unitary operation, such as a CNOT-gate operation $U_{\mathrm{CNOT},jk}=\ket{0}_j {}_j\bra{0}1_k+\ket{1}_j {}_j\bra{1}X_{k}$ on qubits $j$ and~$k$:
\begin{align}
	&\,U_{\mathrm{CNOT},jk}\rho_0 U_{\mathrm{CNOT},jk}^\dag\nonumber\\
	\mapsto&\,(1-\eta_{j,k})U_{\mathrm{CNOT},jk}\rho_0 U_{\mathrm{CNOT},jk}^\dag\nonumber\\
	&\,+\dfrac{\eta_{j,k}}{15}\sum_{1\neq P_{jk}\in\mathcal{P}^{(j,k)}_2}P_{jk}U_{\mathrm{CNOT},jk}\rho_0 U_{\mathrm{CNOT},jk}^\dag P_{jk}\,,
	\label{eq:noisy_CNOT}
\end{align}
where $\eta_{j,k}$ is the error rate from this CNOT operation and $\mathcal{P}^{(j,k)}_2=\{1,X_j,Y_j,Z_j\}\times\{1,X_k,Y_k,Z_k\}$ is the set of two-qubit Pauli operators and the identity for qubits $j$ and~$k$. Hence, the unitary operator $W_1=U_{\mathrm{CNOTs}}R_1$ for the first layer of an $n$-qubit circuit \emph{ansatz} (see Fig.~\ref{fig:ansatz}) consisting of parameter-encoded near-noiseless single-qubit gates $R_1=R^{(1)}_1\otimes R^{(1)}_2\otimes\ldots\otimes R^{(1)}_n$ followed by an array of $n$ noisy CNOT operations $U_{\mathrm{CNOTs}}=U_{\mathrm{CNOT},n\,1}U_{\mathrm{CNOT},n-1\,n}\ldots U_{\mathrm{CNOT},2\,3}U_{\mathrm{CNOT},1\,2}$ gives the noisy state $\rho^{(1)}_{\eta}=W_1\ket{\rvec{0}}(1-\eta_1)\bra{\rvec{0}}W_1^\dag+\eta_1\rho^{(1)}_\mathrm{err}$, with $1-\eta_1=\prod^n_{j=1}(1-\eta_{j,\!\!\!\mod\!\!(j,n)+1})$. It follows that the noisy version of an $L$-layered \emph{ansatz} state defined by $U_{\rvec{\theta}}=W_LW_{L-1}\ldots W_2 W_1$ is given by
\begin{align}
	\rho^{(L)}_{\rvec{\theta},\eta}=&\,U_{\rvec{\theta}}\ket{\rvec{0}}(1-\eta)\bra{\rvec{0}}U_{\rvec{\theta}}^\dag+\eta\rho^{(L)}_\mathrm{err}(\rvec{\theta},\eta)\,,\nonumber\\
	\eta=&\,1-\prod^L_{l=1}\prod^n_{j=1}\left[1-\eta^{(l)}_{j,\!\!\!\!\!\mod\!\!(j,n)+1}\right]\,.
	\label{eq:noisy_ansatz}
\end{align} 
Specifically, when $\eta^{(l)}_{j,k}=\eta_0$ are equal, \mbox{$\eta=1-(1-\eta_0)^{nL}$}. For small $\eta_0$, we consequently find that $\eta\cong nL\eta_0$. 

In all numerical simulations, we shall consider noisy CNOT gates that bring about \eqref{eq:noisy_CNOT} and \eqref{eq:noisy_ansatz} with a constant error rate $\eta_0$ per CNOT gate. All single-qubit gates are always taken to have unit fidelity for granted. The corresponding figures of merit are still the MSEs, but this time, the \emph{estimators} are noisy, whereas the \emph{true} components are not. The MSE definitions are otherwise similar to those in \eqref{eq:MSE_noiseless}.

\begin{figure}[t]
	\centering
	\includegraphics[width=\columnwidth]{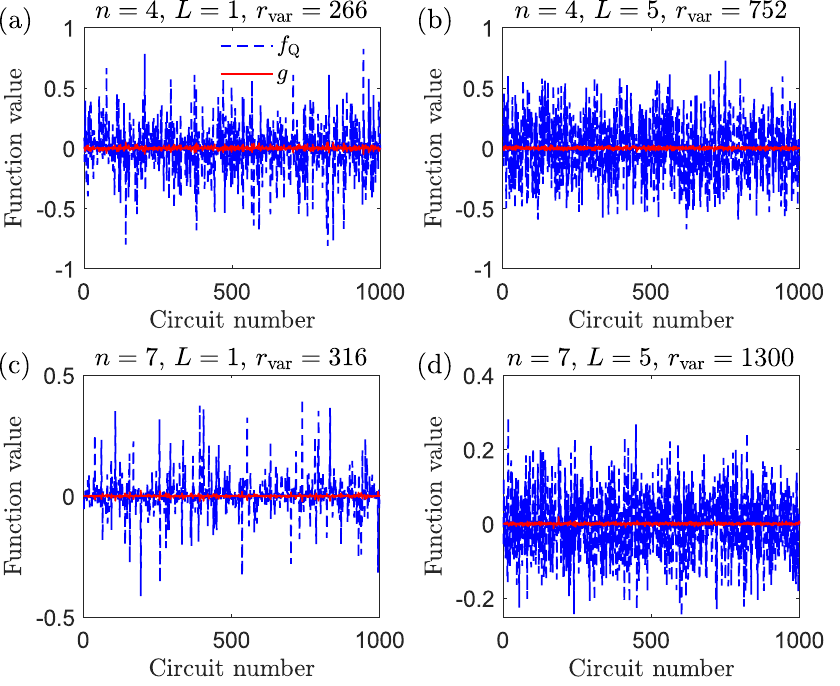}
	\caption{\label{fig:f_g}Distributions of $\fQ$ and $g$ in $f_{\mathrm{Q},\eta}$ for PEPQCs with (a,b)~$n=4$ and (c,d)~$n=7$ qubits over 1000 sets of randomly-generated PEPQC parameters (Haar-distributed single-qubit unitary rotations) in each figure panel. The ratio $r_\mathrm{var}=\VAR{\rvec{\theta}}{\fQ}/\VAR{\rvec{\theta}}{g}$ is given in every panel. The CNOT depolarizing error rate is set at $\eta_0=0.05$ and the overall error rates $\eta=1-(1-\eta_0)^{nL}$ are (a)~0.185, (b)~0.642, (c)~0.302 and (d)~0.834. The respective observables are $O=X_1Y_2Z_3X_4$ and $O=X_1Y_2Z_3X_4Y_5Z_6X_7$.} 
\end{figure}

\section{Result~1: Advantages of using numerical estimators optimized for noiseless circuits}
\label{sec:res1}

\subsection{Noisy MSEs of numerical estimators}
\label{subsec:noisy_MSE}
For the sole purpose of acquiring an analytical understanding of the performance of numerical schemes on noisy circuit, we first assume that $\rho_\mathrm{err}^{(L)}(\rvec{\theta},\eta)=\rho_\mathrm{err}^{(L)}(\eta)$ does not depend on the circuit parameters $\rvec{\theta}$. This implies that the noisy function
\begin{equation}
	{\fQ}_\eta(\rvec{\theta})=(1-\eta)\fQ(\rvec{\theta})+\eta g
	\label{eq:constant_err_term}
\end{equation}
is a sum of the noiseless $\fQ(\rvec{\theta})$ and some constant term $g=\tr{\rho^{(L)}_\mathrm{err}(\eta)\,O}$ that is independent of~$\rvec{\theta}$. Note that $-1\leq g\leq1$ still generally depends on $\eta$, but we shall suppress this dependence for notational simplicity as it shall be irrelevant for subsequent discussions unless otherwise required. 

Figure~\ref{fig:f_g} illustrates that for PEPQCs (that is, all $R_k^{(l)}$ are products of encoded Pauli gates) with the circuit noise channel of uniform error rate $\eta_0$ described in Sec.~\ref{sec:noisy_QC}, which shall be the noise channel of choice for all subsequent simulations, the distribution of $g$ in $\rvec{\theta}$ is generally much flatter than that of $\fQ(\rvec{\theta})$. The constant-$g$ (in $\rvec{\theta}$) assumption thus serves as a reasonable approximation for such a physically-motivated noise model. This approximation is slightly elaborated in Appendix~\ref{app:randPauli_dist_fg}.

The second assumption that we shall make to facilitate the analysis is that all $\rvec{\theta}$ averages are well-approximated by averages over unitary two-designs. This means that the first and second moments of $U_{\rvec{\theta}}$ coincide with the Haar measure of the unitary group~\cite{Puchala_Z._Symbolic_2017,Mele:2023introduction}. This is a rather good approximation as broad classes of circuits with polynomial and logarithmic circuit depths are approximately unitary two-designs~\cite{harrow_random_2009,Dankert:2009exact,Cleve:2016near-linear}. Additionally, we append the implicit technical caveat that \emph{all} gradient and Hessian components are evaluated at parameters whose encoded Pauli gates are each sandwiched by two-design circuit unitary operators. This shall be coined the \emph{two-design sandwich}~(TDS) condition, which is satisfied for the majority of gradient and Hessian components in the bulk of a sufficiently deep circuit.

When $O$ is a traceless Pauli observable, this two-design framework, along with the TDS condition, permits one to obtain PEPQC-based identities that are independent of $(\mu,l)$~\cite{Teo:2023optimized}:
\begin{align}
	\left<\fQ\right>=&\,0\,,\nonumber\\
	\left<\fQ^2\right>=&\,\dfrac{1}{d+1}\,,\nonumber\\
	\left<|\partial\fQ|^2\right>=\left<|\partial\partial\fQ|^2\right>=&\,\dfrac{d^2}{2(d+1)(d^2-1)}\,\,\,\quad(\mathrm{TDS})\,,\nonumber\\
	\left<|\partial\partial'\fQ|^2\right>=&\,\dfrac{d^4}{4(d+1)(d^2-1)^2}\quad(\mathrm{TDS})\,.\label{eq:two-design_idents}
\end{align}

To summarize, in order to acquire some analytical understanding of gradient and Hessian estimation with noisy quantum circuits (in terms of MSE expressions and lemmas), we make the following two main assumptions that are physically reasonable:
\begin{enumerate}
	\item The noise channel generates error terms that are independent of the circuit parameters~$\rvec{\theta}$~(constant $g$ in $\rvec{\theta}$).
	\item The noiseless $U_{\rvec{\theta}}$ is a unitary two-designs, where gradient and Hessian components of its parameters conform to the TDS condition such that~\eqref{eq:two-design_idents} hold.
\end{enumerate}
Under these two assumptions, for traceless Pauli observables, we arrive at the following \emph{exact} MSEs for~FD,
\begin{align}
	\MSE{\mathrm{FD}}{\partial \fQ}=&\,\dfrac{4}{N_\mathrm{T}\epsilon^2}\left[1-(1-\eta)^2\left<\fQ^2\right>-\eta^2g^2\right]\nonumber\\
	&\,+[1-(1-\eta)\,\sinc(\epsilon/2)]^2\left<|\partial\fQ|^2\right>,\nonumber\\
	\MSE{\mathrm{FD}}{\partial\partial\fQ}=&\,\dfrac{18}{N_\mathrm{T}\epsilon^4}\left[1-(1-\eta)^2\left<\fQ^2\right>-\eta^2g^2\right]\nonumber\\
	&\,+\{1-(1-\eta)\,[\sinc(\epsilon/2)]^2\}^2\left<|\partial\partial\fQ|^2\right>,\nonumber\\
	\MSE{\mathrm{FD}}{\partial\partial'\fQ}=&\,\dfrac{16}{N_\mathrm{T}\epsilon^4}\left[1-(1-\eta)^2\left<\fQ^2\right>-\eta^2g^2\right]\nonumber\\
	&\,+\{1-(1-\eta)\,[\sinc(\epsilon/2)]^2\}^2\left<|\partial\partial'\fQ|^2\right>,\label{eq:MSE_FD}
\end{align}
and those for~SPS,
\begin{align}	
	\MSE{\mathrm{SPS}}{\partial \fQ}=&\,\dfrac{\lambda^2}{N_\mathrm{T}}\left[1-(1-\eta)^2\left<\fQ^2\right>-\eta^2g^2\right]\nonumber\\
	&\,+[1-(1-\eta)\,\lambda]^2\left<|\partial\fQ|^2\right>,\nonumber\\
	\MSE{\mathrm{SPS}}{\partial\partial\fQ}=&\,\dfrac{9\lambda^2}{8N_\mathrm{T}}\left[1-(1-\eta)^2\left<\fQ^2\right>-\eta^2g^2\right]\nonumber\\
	&\,+[1-(1-\eta)\,\lambda]^2\left<|\partial\partial\fQ|^2\right>,\nonumber\\
	\MSE{\mathrm{SPS}}{\partial\partial'\fQ}=&\,\dfrac{\lambda^2}{N_\mathrm{T}}\left[1-(1-\eta)^2\left<\fQ^2\right>-\eta^2g^2\right]\nonumber\\
	&\,+[1-(1-\eta)\,\lambda]^2\left<|\partial\partial'\fQ|^2\right>.
	\label{eq:MSE_SPS}
\end{align}
Here, $N_\mathrm{T}$ is the \emph{total sampling-copy number} for estimating the corresponding components. If $N$ copies are used to estimate the function $\fQ$, then $N_\mathrm{T}=2N$ for gradient-component estimation, and $N_\mathrm{T}=3N$ and $4N$ respectively for the diagonal and off-diagonal Hessian-component estimation. Setting $\lambda=1$ gives us the MSEs for PS. The derivation of \eqref{eq:MSE_FD} and~\eqref{eq:MSE_SPS} may be found in Appendix~\ref{app:deriv_MSEs}.

\begin{figure*}[t]
	\centering
	\includegraphics[width=1.8\columnwidth]{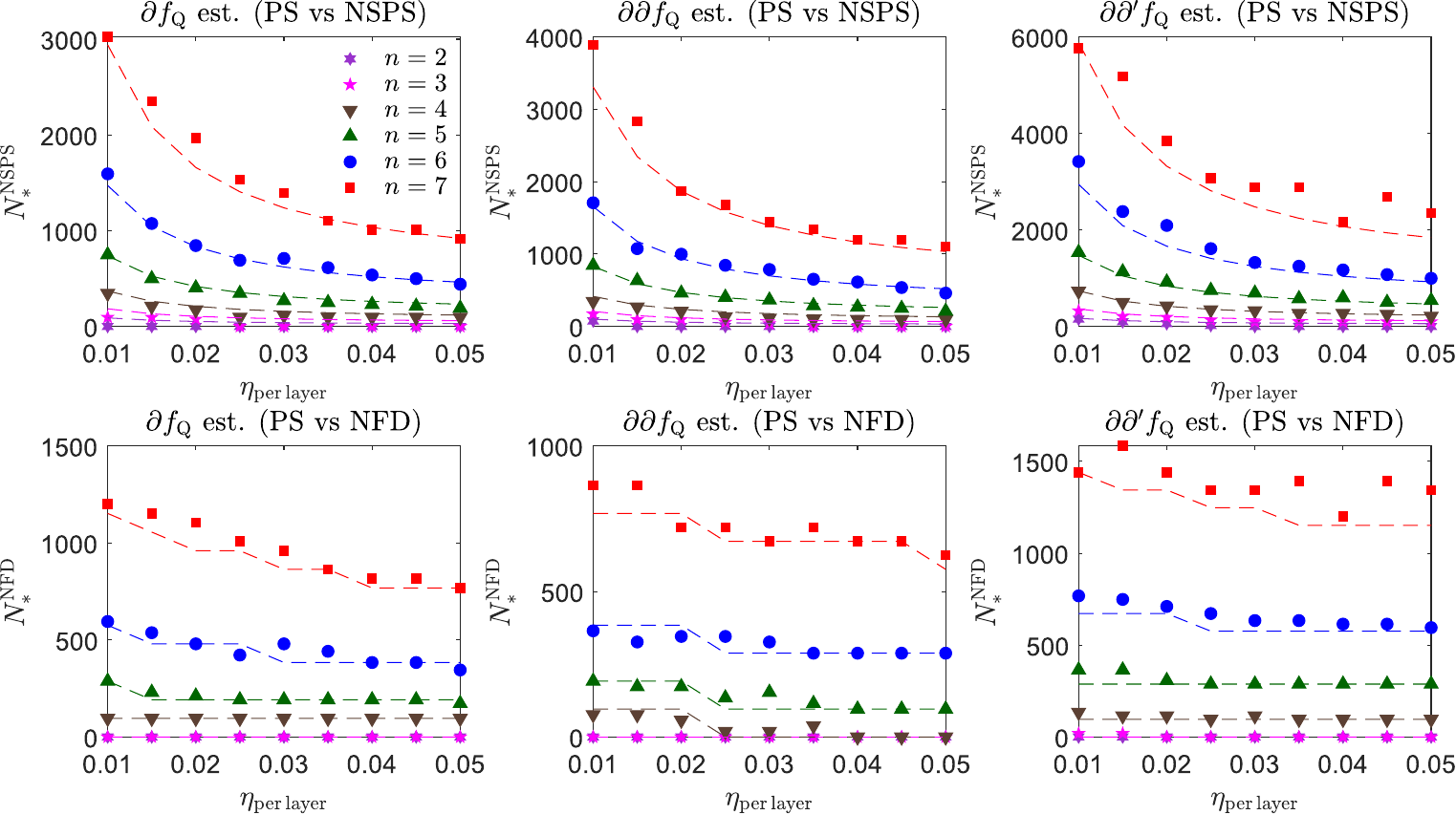}
	\caption{\label{fig:N_star}Monte~Carlo simulations generating the respective $N_*$ behaviors in gradient and Hessian estimations with $\eta_{\text{per\,layer}}$ for $L=5$ in the regime of small $\eta_\text{per layer}$. Noisy CNOT gates as in \eqref{eq:noisy_CNOT} is considered here. The dashed curves in the top figures represent the explicit $N^{\mathrm{NSPS}}_*$ expressions for $g=0$ in \eqref{eq:SPS_N_star_full}, whereas those in the bottom figure trace all $N^{\mathrm{NFD}}_*$ values by numerically solving $\MSE{\mathrm{NFD}}{\cdot}=\MSE{\mathrm{PS}}{\cdot}$ for $N_\mathrm{T}\equiv N^{\mathrm{NFD}}_*$ when $g=0$, all derived based on the two main assumptions. The simulation markers are obtained from MSEs averaged over 1000 sets of random PEPQC parameters (Haar-distributed single-qubit unitary rotations) and 1000 sampling experiments per PEPQC parameter set. We discretize the range of $N_\mathrm{T}$ in steps of 96 copies so that the minimum division is sufficiently large to be distributed to 2, 3 and 4 sampled functions for defining gradient and Hessian estimators as per Sec.~\ref{subsec:noisy_MSE}. We pick $\mu=1$ and $l=2$ to specify the location of the gradient and diagonal Hessian components considered in this figure, and $\mu=1$, $\mu'=2$ and $l=l'=2$ for the off-diagonal Hessian component. As examples, all evaluated gradient and Hessian circuit parameters are encoded onto to the Pauli $Y$-gate. The exponentially increasing trend of $N_*$ with $n$ is numerically evident. The respective observables~$O$ for different $n$ are cyclic repetitions of $X$, $Y$ and $Z$ in this order for every qubit following Fig.~\ref{fig:f_g}.} 
\end{figure*}

\subsection{Naively-optimized numerical schemes and their estimation advantages}
\label{subsec:naive_numer}

It is now possible to compare the performances of SPS and PS in terms of gradient and Hessian estimation. In the noiseless case ($\eta=0$), it is a straightforward matter to deduce~\cite{Teo:2023optimized} that SPS estimators with $\lambda$ taking the following optimal values 
\begin{align}
	\lambda_{\mathrm{opt}}=&\,\dfrac{dN_\mathrm{T}}{2d^2 + dN_\mathrm{T}-2}\leq1\quad\!\qquad(\text{$\partial \fQ$ estimation})\,,\nonumber\\
	\lambda_{\mathrm{opt}}=&\,\dfrac{4dN_\mathrm{T}}{9d^2 + 4dN_\mathrm{T}-9}\leq1\,\,\qquad(\text{$\partial\partial \fQ$ estimation})\,,\nonumber\\
	\lambda_{\mathrm{opt}}=&\,\dfrac{d^3N_\mathrm{T}}{4(d^2-1)^2 + d^3N_\mathrm{T}}\leq1\,\quad(\text{$\partial\partial' \fQ$ estimation})\,,
	\label{eq:SPS_lbd_opt}
\end{align}
will minimize their respective MSEs, and these optimized SPS estimators \emph{always} offer smaller MSEs than those of PS regardless of the value of $N_\mathrm{T}$. For $0<\eta\leq1$, when the \emph{same} $\lambda_{\mathrm{opt}}$s in \eqref{eq:SPS_lbd_opt} are naively used in spite of the presence of noise, a smaller \emph{naively-optimized SPS}~(NSPS) MSE can still be expected when $N_\mathrm{T}$ is below some critical $N_*$. So, a larger $N_*$ signifies a more advantageous numerical scheme over PS. One can directly calculate $N_*$ by setting $\MSE{\mathrm{NSPS}}{\cdot}-\MSE{\mathrm{PS}}{\cdot}=0$, which is generally a complicated function of $\eta$, $g$ and $d$. The explicit expressions of $N_*$ when $g=0$, which is almost exact for our noise model from Fig.~\ref{fig:f_g}, are given in~\eqref{eq:SPS_N_star_full}.

To get a handle on the behavior of $N_*$ in relation to $d$ and~$\eta$, we may further consider small $\eta$ values such that all MSEs of the NSPS schemes may be expanded up to first order in $\eta$. In this small-$\eta$ limit, the leading term $\rho^{(L)}_\mathrm{err}(\eta)$ is also a constant in~$\eta$. The resulting $N_*=N_*^{\mathrm{NSPS}}$ values then read
\begin{align}
	N^{\mathrm{NSPS}}_{*}=&\,\dfrac{(d^2-1)}{d\,\eta}\quad\,\,\qquad(\text{$\partial \fQ$ estimation})\,,\nonumber\\
	N^{\mathrm{NSPS}}_{*}=&\,\dfrac{9(d^2-1)}{8\,d\,\eta}\,\,\,\,\,\qquad(\text{$\partial\partial \fQ$ estimation})\,,\nonumber\\
	N^{\mathrm{NSPS}}_{*}=&\,\dfrac{2(d^2-1)^2}{d^3\eta}\,\,\qquad(\text{$\partial\partial' \fQ$ estimation})\,.
	\label{eq:SPS_N_star}
\end{align}

Therefore, we have the following lemma:
\begin{lemma}
	\label{lem:NSPS_PS}
	\textbf{Gradient- and Hessian-estimation performance advantage of NSPS over PS}---For an $n$-qubit two-design PEPQC that satisfies the TDS condition and any noise channel with a $\rvec{\theta}$-independent error term leading to Eq.~\eqref{eq:constant_err_term}, NSPS outperforms PS if \mbox{$N_\mathrm{T}<N_*=N^{\mathrm{NSPS}}_*\sim O(2^n/\eta)$.}
\end{lemma}
This result tells us that when the PEPQC is noisy, even without knowing \emph{anything} about the noise channel (such as the error rate $\eta$), the NSPS estimators, that is those in \eqref{eq:SPS_noiseless} evaluated with $\lambda=\lambda_\mathrm{opt}$ in \eqref{eq:SPS_lbd_opt}, can still give more accurate estimation than PS estimators when the sampling-copy number is limited. For circuits of very many qubits, owing to the influence of barren plateaus for universal PEPQC \emph{ansat{\"z}e}, Lemma~\ref{lem:NSPS_PS} informs us that these NSPS estimators are still the more accurate ones for very large copy numbers.

We may also naively optimize the FD estimators in much the same way as we did the NSPS estimators. Accordingly, we may consider the noiseless FD MSEs by setting $\eta=0$ in all of~\eqref{eq:MSE_FD} and minimize each of them over $\epsilon$. Since $\epsilon$ enters the MSEs in a transcendental fashion, this minimization is done numerically. The optimal $\epsilon_\mathrm{opt}$ is then a function of $N_\mathrm{T}$ and $d$, much like $\lambda_\mathrm{opt}$ in \eqref{eq:SPS_lbd_opt}. If we now use these $\epsilon_\mathrm{opt}$ to define the FD estimators for noisy circuits, we then have the analogous \emph{naively-optimized~FD} estimators~(NFD) and $N_*=N^{\mathrm{NSPS}}_*$ for them can similarly be found by noting the instant \mbox{$\MSE{\mathrm{NFD}}{\cdot}=\MSE{\mathrm{PS}}{\cdot}$}. 

In Fig.~\ref{fig:N_star}, we compare the values of $N_*$ for NFD and NSPS schemes by simulating PEPQCs possessing noisy CNOT gates with the channel action in~\eqref{eq:noisy_CNOT}. We do this for different number of qubits $n$ given a~\emph{fixed error rate per layer}---$\eta_\text{per layer}$. This allows us to compare MSE performances for various $n$ fairly in an \emph{ansatz}-free manner since the two-qubit gate count that scales with $n$ is absorbed into this ``error rate per layer'' definition. For the circuit \emph{ansatz} described in Sec.~\ref{sec:noisy_QC} which we are considering, $\eta_\text{per layer}\equiv 1-(1-\eta_0)^{n}$. This is also $\eta_\text{per layer}\cong n\eta_0$ when $\eta_0$ is small. The total error rate is hence $\eta\cong\eta_\text{per layer}L$. Overall, naively ignoring noise by using the NSPS schemes can still achieve lower MSEs in contrast to~PS. Compared to NFD, this happens for larger sampling-copy number ranges (or a larger $N_*$).

Nonetheless, both NSPS and NFD schemes, which are optimally tuned for noiseless quantum circuits, will evidently introduce \emph{noise biases} (or MSE biases) even when $N_\mathrm{T}\rightarrow\infty$ if they are employed for noisy circuits. This is a consequence of optimizing over $\epsilon$ and $\lambda$ by ignoring the presence of noise. As these noise biases are permanent systematic errors that cannot be eliminated even when $N_\mathrm{T}=\infty$, they are \emph{not to be confused} with statistical biases of the numerical estimators as explained in Sec.~\ref{sec:numer_analytic}, which, when correctly optimized, can help eliminate noise biases as a matter of fact. In the next section, we discuss how these parameters may be properly tuned for noisy quantum circuits so that the resulting optimized statistical biases actually reduce noise biases.


\section{Result~2: Heuristic error-mitigation for numerical gradient and Hessian estimations}
\label{sec:heuristic}

In the previous section, we found that numerical gradient and Hessian estimation schemes, when optimized under the negligence of noise in PEPQCs, can still give more accurate estimators for a given total sampling-copy number $N_\mathrm{T}$ value no larger than some critical value $N_*$. There is, however, one more procedure we may carry out to further improve the estimation quality of numerical estimators with noisy NISQ circuits. In this section, we shall discuss a simple error-mitigation strategy that can be carried out if the error rate~$\eta$ of the overall noise channel is known \emph{a priori}, which may be acquired through an initial device and channel calibration tests. Otherwise, no other explicit knowledge about the type of noise channel acting on a PEPQC is necessary.

Very briefly, this error-mitigation technique involves making a few heuristic assumptions such that the MSE expressions in~\eqref{eq:MSE_FD} and \eqref{eq:MSE_SPS} are valid, followed by eliminating the dependence of $g$ by considering the upper bounds of these MSE expressions, and finally minimizing these upper bounds to obtain the respective optimal parameters to be used for defining the heuristically optimized numerical estimators given a specific $\eta$. As this technique does not require an accurate noise-channel superoperator description, such an error-mitigation scheme is appealingly feasible in real experimental situations, since a full characterization of $16^n$ real parameters of the noise-channel map is generally impractical for large~$n$.

Let us now describe this error-mitigation protocol with more detail. As the primary step, we make the assumptions listed in Sec.~\ref{subsec:noisy_MSE}, which we repeat here once more: namely that (1)~the noise channel generates error terms (or $g$) that are constant in~$\rvec{\theta}$, and (2)~all $U_{\rvec{\theta}}$ is a unitary two-design in which gradient and Hessian components satisfy the TDS condition, which leads to the MSE expressions in \eqref{eq:MSE_FD} and \eqref{eq:MSE_SPS}. In order to execute this error-mitigation strategy without the knowledge about the noise channel or $g$, the second step we shall take is to consider the upper bounds of the MSE expressions, which are obtained by simply removing ``$\eta^2g^2$'' on the right-hand sides of \eqref{eq:MSE_FD} and~\eqref{eq:MSE_SPS}. After minimizing the MSE upper bounds, the $\lambda_{\mathrm{opt},\eta}$ parameters characterizing the \emph{heuristically-optimized SPS} scheme~(HSPS) are
\begin{align}
	\lambda_{\mathrm{opt},\eta}=&\,\dfrac{dN_\mathrm{T}(1-\eta)}{2d^2 + dN_\mathrm{T}-2+\eta(2-\eta)(2d-dN_\mathrm{T}-\frac{2}{d})}\nonumber\\
	&\,\qquad\qquad\qquad\qquad\qquad\qquad\,\,\,\,\,\,(\text{$\partial \fQ$ estimation})\,,\nonumber\\
	\lambda_{\mathrm{opt},\eta}=&\,\dfrac{4dN_\mathrm{T}(1-\eta)}{9d^2 + 4dN_\mathrm{T}-9+\eta(2-\eta)(9d-4dN_\mathrm{T}-\frac{9}{d})}\nonumber\\
	&\,\qquad\qquad\qquad\qquad\qquad\qquad\,\,\,(\text{$\partial\partial \fQ$ estimation})\,,\nonumber\\
	\lambda_{\mathrm{opt},\eta}=&\,\dfrac{d^3N_\mathrm{T}(1-\eta)}{4(d^2-1)^2 + d^3N_\mathrm{T}+\eta(2-\eta)[\frac{4}{d}(d^2-1)^2-d^3N_\mathrm{T}]}\nonumber\\
	&\,\qquad\qquad\qquad\qquad\qquad\qquad\,\,\,(\text{$\partial\partial' \fQ$ estimation})\,.
	\label{eq:SPS_lbd_opt_eta}
\end{align}

\begin{figure*}[t]
	\centering
	\includegraphics[width=1.8\columnwidth]{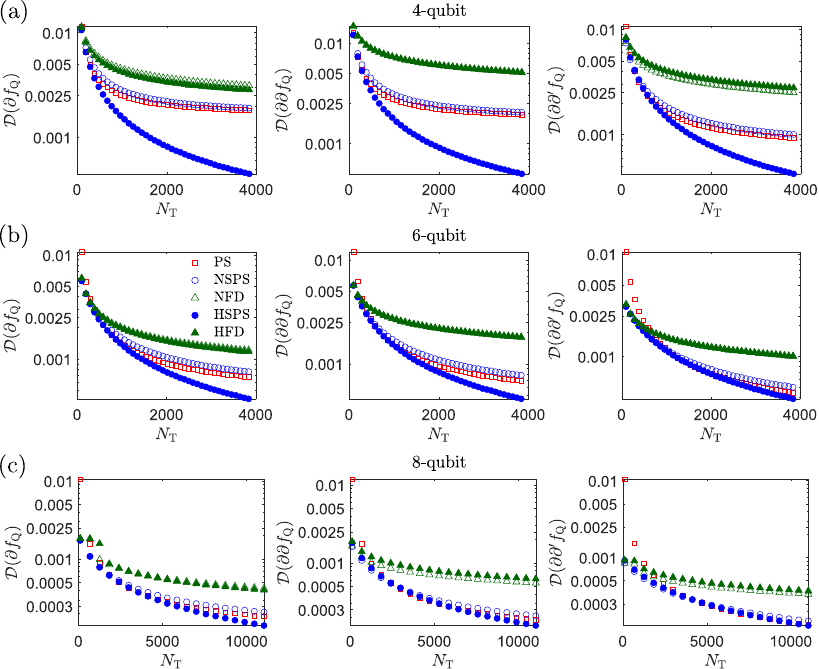}
	\caption{\label{fig:heuristic}Monte~Carlo simulations comparing the naive~(NFD, NSPS) and heuristically optimized~(HFD, HSPS) numerical gradient and Hessian estimation schemes with (a)~four-qubit, (b)~six-qubit and (c)~eight-qubit PEPQCs for $\eta_\text{per layer}=0.05$ and $L=5$ circuit layers~[$\eta=1-(1-\eta_\text{per layer})^L=0.226\cong0.25$]. The gradient and diagonal Hessian components are specified by $\mu=1$ and $l=2$, and the off-diagonal Hessian component by $\mu=1$, $\mu'=2$ and $l=l'=2$, with evaluated gradient and Hessian circuit parameters encoded to the $Y$-gate. All MSEs are averaged over 500 sets of random PEPQC parameters (Haar-distributed single-qubit unitary rotations) and 500 sampling experiments per PEPQC parameter set. In these plots, we observe that while HSPS significantly improves the MSEs with respect to NSPS and PS, there is almost no visible differences between HFD and NFD estimators even with the logarithmic-scale plots presented here. The respective observables~$O$ for the different $n$ values are cyclic repetitions of $X$, $Y$ and $Z$ in this order for every qubit, as in the captions of Figs.~\ref{fig:f_g} and~\ref{fig:N_star}.} 
\end{figure*}

When $\eta$ is known \emph{a priori}, such an error-mitigation strategy of minimizing MSE upper bounds over the parameters that characterize numerical estimators can reduce their noise biases significantly. Furthermore, the simplicity of the $\lambda$-dependence in all SPS MSEs allows us to acquire a quantitative understanding of these noise biases.
\begin{lemma}
	\label{lemm:hsps}
	\textbf{Gradient- and Hessian-estimation performance advantage of HSPS over NSPS and PS}---Suppose an $n$-qubit two-design PEPQC satisfying the TDS condition is subjected to a noise channel of a fixed error rate $\eta>0$ and $\rvec{\theta}$-independent error term leading to Eq.~\eqref{eq:constant_err_term}. Then when \mbox{$N_\mathrm{T}\rightarrow\infty$}, the NSPS and PS estimators give nonzero MSEs, which are respectively $\left<|\partial\fQ|^2\right>\eta^2$ for gradient estimation, $\left<|\partial\partial\fQ|^2\right>\eta^2$ and $\left<|\partial\partial'\fQ|^2\right>\eta^2$ respectively for diagonal and off-diagonal Hessian estimation. On the other hand, the MSEs of HSPS estimators asymptotically approach zero.
\end{lemma}

This heuristic protocol may also apply to the FD scheme, which would then yield the \emph{heuristically-optimized FD} scheme~(HFD). However, as in the NFD estimation protocol in Sec.~\ref{subsec:naive_numer}, the $\epsilon_{\mathrm{opt},\eta}$s for all gradient and Hessian estimations have no closed forms and should be obtained by numerically minimizing the upper bounds of the MSEs in~\eqref{eq:MSE_FD}. Unfortunately, it turns out that the HFD estimators acquired this way are just as noisily biased as the NFD ones. This is equivalently encapsulated in the next lemma:
\begin{lemma}
	\label{lemm:hfd}
	\textbf{HFD and NFD schemes are asymptotically noisy}---Suppose an $n$-qubit two-design PEPQC satisfying the TDS condition is subjected to a noise channel of a fixed error rate $\eta>0$ and $\rvec{\theta}$-independent error term leading to Eq.~\eqref{eq:constant_err_term}. Then when $N_\mathrm{T}\rightarrow\infty$, the NFD and HFD estimators give nonzero MSEs, which are respectively $\left<|\partial\fQ|^2\right>\eta^2$ for gradient estimation, $\left<|\partial\partial\fQ|^2\right>\eta^2$ and $\left<|\partial\partial'\fQ|^2\right>\eta^2$ respectively for diagonal and off-diagonal Hessian estimation.
\end{lemma}
The interested Reader may refer to Appendix~\ref{app:biases} for the simple arguments leading to Lemmas~\ref{lemm:hsps} and~\ref{lemm:hfd}.

One can intuitively understand the reasons behind these two lemmas by observing that for the HSPS estimators to completely eliminate noise biases, in the limit of large $N_\mathrm{T}$, we expect the condition $(1-\eta)\lambda_{\mathrm{opt},\eta}=1$ to hold in order for the approximation error to approach zero, which means that $\lambda_{\mathrm{opt},\eta}>1$, consistent with the large-$N_\mathrm{T}$ versions of~\eqref{eq:SPS_lbd_opt_eta}. For the HFD estimators, demanding an asymptotically vanishing approximation error would entail the obedience of the analogous condition $(1-\eta)\sinc(\epsilon_{\mathrm{opt},\eta}/2)=1$, which can never happen for \emph{any} $\eta>0$ since the sinc function is always less than or equal to~1. The only way out is $\eta=0$, where we recover the textbook asymptotic limit $\epsilon_{\mathrm{opt},0}=\epsilon_{\mathrm{opt}}\rightarrow0$.

Figure~\ref{fig:heuristic} compares the NSPS and NFD schemes with their heuristically optimized counterparts HSPS and HFD using known $\eta_0$ or $\eta_\text{per layer}$, where the quantum circuit \emph{ansatz} comprises layers of single-qubit gates followed by a complete array of noisy CNOT gates. The zero-noise-bias property of HSPS for constant-$g$ noise channels as in Lemma~\ref{lemm:hsps} carries over to other more general noise models, such as the one governed by \eqref{eq:noisy_CNOT}. However, because of the noise biases that persist in HFD estimators, such an error mitigation does improve the MSE. Moreover, the close competition between NFD and HFD and the proximity of their asymptotic noise biases may result in NFD estimators giving lower MSEs than HFD estimators. This is, however, allowed as only the upper bounds of the MSE are minimized to obtain the heuristic schemes, so there is not guarantee for HFD to always do better than NFD. Meanwhile, HSPS estimators would always outperform NSPS and PS for sufficiently large $N_\mathrm{T}$ if $g$ is roughly a constant in~$\rvec{\theta}$.

These findings suggest that for quantum circuits of sufficient depth and general noisy channels of a \emph{known} error rate, whose error terms are approximately constants in the circuit parameters, HSPS is an advantageous gradient and Hessian estimation scheme that can significantly reduce asymptotic noise biases for any given finite number of sampling copies. As the qubit number $n\rightarrow\infty$, the asymptotic noise biases of all estimators eventually go to zero, but so are the gradient and Hessian magnitudes anyway, which is a signature of the barren-plateau problem. In this detrimental regime, \emph{no} estimation is feasible.

\section{Conclusion}

The results of this work and those presented in the prequel article revolve around the use of numerical methods (such as finite-difference and scaled parameter-shift rule) to estimate the gradient and Hessian of quantum-circuit functions in variational quantum algorithms. With hypothetical noiseless circuits, it is known from the prequel article that a proper optimization of statistical biases in numerical estimators can achieve lower mean-squared errors (or synonymously sampling errors) than analytical ones~(namely estimators derived from the unscaled parameter-shift rule). The key point is that sampling is required in variational algorithms, and the additional parameter degree of freedom in numerical estimators permits us to achieve optimal mean-squared errors that drop exponentially with the number of qubits, commensurate with the exponentially-decaying gradient and Hessian magnitudes as a result of barren plateaus occurring in deep universal quantum circuits. This possibility is absent in analytical estimators.

With realistic noisy quantum circuits, we have shown here that when one uses numerical estimators that are optimized for noiseless circuits to estimate gradient and Hessian components of noisy circuit functions, there exist nonzero sampling-copy-number regimes where these so-called naively-optimized numerical estimators can still achieve lower mean-squared errors than analytical estimators. Under two physically realistic assumptions on both the quantum-circuit structure and noise channel as stated in Secs.~\ref{subsec:noisy_MSE}, we explicitly show that the scaled parameter-shift estimators outperform the corresponding unscaled analytical ones within a sampling-copy-number range that increases exponentially with the qubit number, and also increases reciprocally with the total noise-channel error rate. These properties are carried over to practical channels modeling noisy two-qubit gates on a layered circuit \emph{ansatz}.

These naively-optimized numerical estimators have innate noise biases that do not result in faithful gradient and Hessian estimation even when the sampling-copy number becomes infinity, as they are strictly not meant for noisy quantum circuits. To resolve this problem, we proposed an experimentally operational error-mitigation technique that does not require the precise knowledge concerning the type of noise channel acting on the circuit; only the error rate is needed. This technique employs, again, the two assumptions about the circuit and noise channel, and seeks to minimize the mean-squared error upper bounds of numerical schemes to obtain heuristically-optimized estimators that are more compatible with noisy circuits. Indeed, we showed that the heuristically-optimal scaled parameter-shift estimators not only completely eliminate noise biases under noise channels with constant error terms, but also significantly reduce these noise biases when physically realistic circuit noise models are considered. The heuristically-optimized finite-difference estimators, unfortunately, are just as noisily biased as their naively-optimized counterparts, and should be avoided.

The heuristic nature of the error-mitigation procedure introduced in this work originates from the two assumptions about the circuit unitary properties and noise channels. We emphasize, however, that since these assumptions are approximately aligned with moderately-deep quantum circuits and realistic circuit noise channels, the corresponding heuristically-optimal scaled parameter-shift estimators are consequently also relevant and interesting in practical situations.

As in the prequel article, we reiterate here that having a statistically accurate estimation scheme is the first of many important steps towards the goal of trainable quantum circuits, and that this quest is by no means finished with this work. Much more efforts are required in seeking new initialization strategies and more expressive circuit \emph{ans{\"a}tze} to circumvent the barren plateau problem, or, more generally, the concentration-of-measure phenomenon in variational quantum algorithms.

\begin{acknowledgements}
	This work is supported by the National Research Foundation of Korea (NRF) grants funded by the Korea government (Grant nos.~NRF-2020R1A2C1008609, NRF-2020K2A9A1A06102946, RS-2023-00237959 and NRF-2022M3E4A1076099) \emph{via} the Institute of Applied Physics at Seoul National University, and the Brain Korea~21 FOUR Project grant funded by the Korean Ministry of Education.
\end{acknowledgements}

\appendix

\section{Remarks on constant noise-channel error terms}
\label{app:randPauli_dist_fg}

\begin{figure}[h!]
	\centering
	\includegraphics[width=\columnwidth]{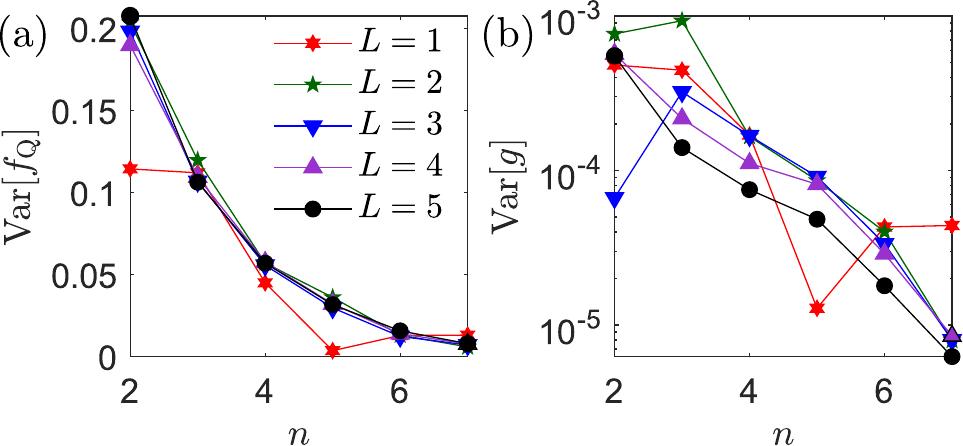}
	\caption{\label{fig:var_n}Plots of (a)~$\VAR{\rvec{\theta}}{\fQ}$ and (b)~$\VAR{\rvec{\theta}}{g}$ against $n$ for $\eta_0=0.05$ and various $L$ values under the CNOT-gate depolarizing channel. All curves are averaged over 1000 random sets of PEPQC parameters (Haar-distributed single-qubit unitary rotations).} 
\end{figure}

Figure~\ref{fig:var_n} illustrates the behaviors of $\VAR{\rvec{\theta}}{\fQ}$ and $\VAR{\rvec{\theta}}{g}$ with respect to $n$ for the CNOT-gate depolarizing channel defined in \eqref{eq:noisy_CNOT} with a uniform error rate $\eta_0$. The decreasing variances over $n$ is an indication of the concentration of measure phenomenon that occurs when the number of free parameters or system dimension tends to infinity~\cite{Barvinok:1997measure,Ledoux:2001concentration,Muller:2011concentration}. The ratio $r_\mathrm{var}$ tends to increase with $n$, especially when $L$ is large.

\begin{figure}[t]
	\centering
	\includegraphics[width=\columnwidth]{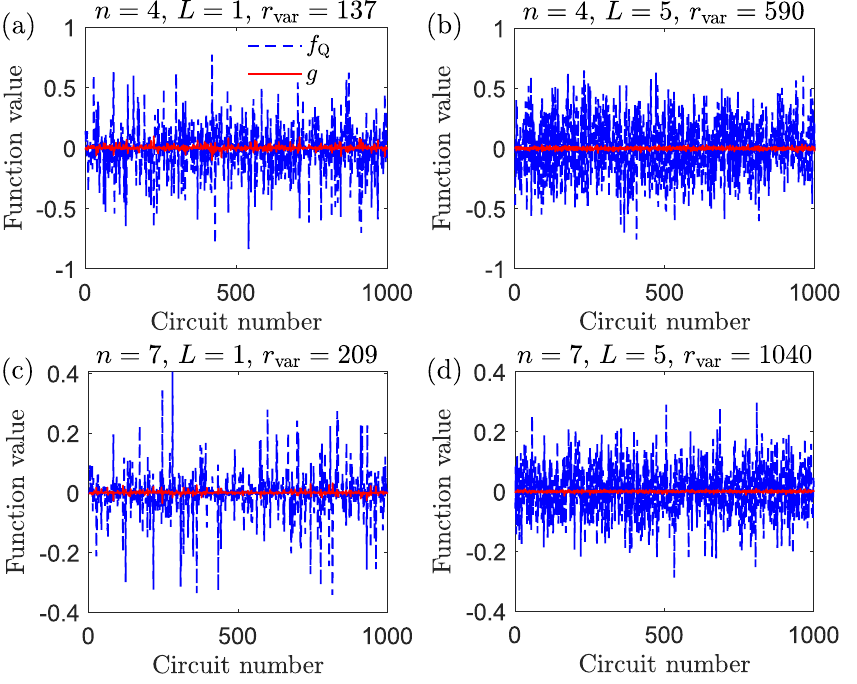}
	\caption{\label{fig:randPauli_f_g}Distributions of $\fQ$ and $g$ in $f_{\mathrm{Q},\eta}$ for PEPQCs with (a,b)~$n=4$ and (c,d)~$n=7$ qubits over 1000 sets of randomly-generated PEPQC parameters (Haar-distributed single-qubit unitary rotations) in each figure panel. The ratio $r_\mathrm{var}=\VAR{\rvec{\theta}}{\fQ}/\VAR{\rvec{\theta}}{g}$ is given in every panel. The Pauli-channel error rate is set at $\eta_0=0.05$ and $\rvec{\eta}_0$ is randomly chosen for each PEPQC parameter set. The overall error rates $\eta=1-(1-\eta_0)^{nL}$ are (a)~0.185, (b)~0.642, (c)~0.302 and (d)~0.834. The respective observables are $O=X_1Y_2Z_3X_4$ and $O=X_1Y_2Z_3X_4Y_5Z_6X_7$ for $n=4$ and 7.} 
\end{figure}

To show that the constant-$g$ assumption is not bad even for general Pauli channels, we also look at the distribution of~$g$ for the case in which the noise-channel map~$\mathcal{E}$ corresponds to a general Pauli channel where the 15-dimensional column $\rvec{\eta}_0\,\widehat{=}\,(\eta_{0,12},\eta_{0,13},\ldots,\eta_{0,44})^\top$ per noisy CNOT gate has all entries that sum to some fixed $\eta_0$. The action is then given by
\begin{align}
	&\,U_{\mathrm{CNOT},jk}\rho_0 U_{\mathrm{CNOT},jk}^\dag\nonumber\\
	\mapsto&\,(1-\eta_0)U_{\mathrm{CNOT},jk}\rho_0 U_{\mathrm{CNOT},jk}^\dag\nonumber\\
	&\,+\sum_{1\neq P_{jk}\in\mathcal{P}^{(j,k)}_2}\eta_{0,jk}P_{jk}U_{\mathrm{CNOT},jk}\rho_0 U_{\mathrm{CNOT},jk}^\dag P_{jk}\,.
	\label{eq:noisy_CNOT_Pauli}
\end{align}
Figure~\ref{fig:randPauli_f_g} shows a very similar characteristic for $g$ in which and that $r_\mathrm{var}$ increases with increasing $n$ and $L$.

\section{Derivations of Eqs.~\eqref{eq:MSE_FD} and \eqref{eq:MSE_SPS}}
\label{app:deriv_MSEs}

We shall present the derivations of $\MSE{\mathrm{FD}}{\partial \fQ}$ and $\MSE{\mathrm{SPS}}{\partial \fQ}$. All other expressions may be obtained in a similar fashion. Starting with the general definition
\begin{equation}
	\MSE{\,\bm{\cdot}\,}{\partial \fQ}=\left<\overline{\left(\widehat{[\partial_{\,\bm{\cdot}\,}]\fQ}_\eta-\partial\fQ\right)^2}\right>\,,
\end{equation}
where $\widehat{[\partial_{\,\bm{\cdot}\,}]\fQ}_\eta$ is the gradient estimator subjected to noise of error rate $\eta$ (the subscripts $\mu$ and $l$ will be dropped in this discussion). Note that
\begin{align}
	\MSE{\,\bm{\cdot}\,}{\partial \fQ}=&\,\underbrace{\left<\overline{\left(\widehat{[\partial_{\,\bm{\cdot}\,}]\fQ}_\eta-[\partial_{\,\bm{\cdot}\,}]{\fQ}_\eta\right)^2}\right>}_{\displaystyle\text{finite-copy error}}\nonumber\\
	&\,+\underbrace{\left<\left([\partial_{\,\bm{\cdot}\,}]{\fQ}_\eta-\partial\fQ\right)^2\right>}_{\displaystyle\text{approximation error}}
	\label{eq:finite_approx_errs}
\end{align}
is a sum of the finite-copy and approximation errors. Gradient estimators are formed by taking the difference between two translated circuit functions and dividing it by a scalar. 

For FD, this scalar is twice the translation according to \eqref{eq:FD_noiseless}, so that the independence of the data collected for each translated function results in
\begin{align}
	&\,\left<\overline{\left(\widehat{[\partial_{\mathrm{FD}}]\fQ}_\eta-[\partial_{\mathrm{FD}}]{\fQ}_\eta\right)^2}\right>\nonumber\\
	=&\,\dfrac{1}{\epsilon^2}\Bigg\{\left<\overline{\left[\widehat{{\fQ}_\eta\left(\theta+\frac{\epsilon}{2}\right)}-{\fQ}_\eta\left(\theta+\frac{\epsilon}{2}\right)\right]^2}\right>\nonumber\\
	&\,+\left<\overline{\left[\widehat{{\fQ}_\eta\left(\theta-\frac{\epsilon}{2}\right)}-{\fQ}_\eta\left(\theta-\frac{\epsilon}{2}\right)\right]^2}\right>\Bigg\}
\end{align}
Since ${\fQ}_\eta(\theta)=\tr{\rho_{\theta,\eta}O}=\sum^{d-1}_{k=0}o_k\opinner{k}{\rho_{\theta,\eta}}{k}$, where $\ket{k}$ is an eigenket of $O$ with eigenvalue~$o_k$, $\widehat{{\fQ}_\eta(\theta)}$ can be defined as an unbiased estimator of ${\fQ}_\eta(\theta)$ inasmuch as
\begin{equation}
	\widehat{{\fQ}_\eta(\theta)}=\sum^{d-1}_{k=0}o_k\nu_{k,\theta,\eta}=\dfrac{1}{N}\sum^{d-1}_{k=0}o_kn_{k,\theta,\eta}\,
\end{equation}
with $\nu_{k,\theta,\eta}\rightarrow p_{k,\theta,\eta}=\opinner{k}{\rho_{\theta,\eta}}{k}$ in the limit of large~$N$, such that $\overline{\widehat{{\fQ}_\eta(\theta)}}={\fQ}_\eta(\theta)$ as $\overline{\nu_{k,\theta,\eta}}= p_{k,\theta,\eta}$. So, using the identity
\begin{equation}
	\overline{\nu_{k,\theta,\eta}\nu_{k',\theta,\eta}}-p_{k,\theta,\eta}p_{k',\theta,\eta}=\dfrac{1}{N}(\delta_{k,k'}p_{k,\theta,\eta}-p_{k,\theta,\eta}p_{k',\theta,\eta})
	\label{eq:multinom}
\end{equation}
for the multinomial distribution, a traceless Pauli observable~$O$ implies that
\begin{align}
	&\,\left<\overline{\left[\widehat{{\fQ}_\eta\left(\theta+\frac{\epsilon}{2}\right)}-{\fQ}_\eta\left(\theta+\frac{\epsilon}{2}\right)\right]^2}\right>\nonumber\\
	=&\,\dfrac{1}{N}\left<\sum^{d-1}_{k,k'=0}o_ko_{k'}(\delta_{k,k'}p_{k,\theta+\epsilon/2,\eta}-p_{k,\theta+\epsilon/2,\eta}p_{k',\theta+\epsilon/2,\eta})\right>\nonumber\\
	=&\,\dfrac{1}{N}\left(1-\left<{\fQ}_\eta\left(\theta+\frac{\epsilon}{2}\right)^2\right>\right)\nonumber\\
	=&\,\dfrac{1}{N}\left[1-(1-\eta)^2\left<\fQ^2\right>-\eta^2g^2\right]\,,
\end{align}
where we made use of the constant-$g$ and two-design assumptions---$\left<\fQ g\right>=\left<\fQ \right>g=0$. The finite-copy error hence reads
\begin{align}
	&\,\left<\overline{\left(\widehat{[\partial_{\mathrm{FD}}]\fQ}_\eta-[\partial_{\mathrm{FD}}]{\fQ}_\eta\right)^2}\right>\nonumber\\
	=&\,\dfrac{2}{N\epsilon^2}\left[1-(1-\eta)^2\left<\fQ^2\right>-\eta^2g^2\right]\,.
\end{align}
The approximation error is straightforward to cope with using the same two assumptions:
\begin{align}
	&\,\left<\left([\partial_{\mathrm{FD}}]{\fQ}_\eta-\partial\fQ\right)^2\right>\nonumber\\
	=&\,\left<\left\{\dfrac{1}{\epsilon}\left[{\fQ}_\eta\left(\theta+\frac{\epsilon}{2}\right)-{\fQ}_\eta\left(\theta-\frac{\epsilon}{2}\right)\right]-\partial\fQ\right\}^2\right>\nonumber\\
	=&\,\left<\left\{\dfrac{1-\eta}{\epsilon}\left[{\fQ}\left(\theta+\frac{\epsilon}{2}\right)-{\fQ}\left(\theta-\frac{\epsilon}{2}\right)\right]-\partial\fQ\right\}^2\right>\nonumber\\
	=&\,\left<\left[(1-\eta)\,\sinc\left(\frac{\epsilon}{2}\right)\partial\fQ-\partial\fQ\right]^2\right>\nonumber\\
	=&\,\left[1-(1-\eta)\,\sinc\left(\frac{\epsilon}{2}\right)\right]^2\left<|\partial\fQ|^2\right>\,.
\end{align}
What is left the assignment $N_\mathrm{T}=2N$.

By the same token, the SPS finite-copy error is given by
\begin{align}
	&\,\left<\overline{\left(\widehat{[\partial_{\mathrm{SPS}}]\fQ}_\eta-[\partial_{\mathrm{SPS}}]{\fQ}_\eta\right)^2}\right>\nonumber\\
	=&\,\dfrac{\lambda^2}{4}\Bigg\{\left<\overline{\left[\widehat{{\fQ}_\eta\left(\theta+\frac{\pi}{2}\right)}-{\fQ}_\eta\left(\theta+\frac{\pi}{2}\right)\right]^2}\right>\nonumber\\
	&\,+\left<\overline{\left[\widehat{{\fQ}_\eta\left(\theta-\frac{\pi}{2}\right)}-{\fQ}_\eta\left(\theta-\frac{\pi}{2}\right)\right]^2}\right>\Bigg\}\nonumber\\
	=&\,\dfrac{\lambda^2}{2N}\left[1-(1-\eta)^2\left<\fQ^2\right>-\eta^2g^2\right]\,.
\end{align}
Likewise, its approximation error is straightforwardly acquired through these steps:
\begin{align}
	&\,\left<\left([\partial_{\mathrm{SPS}}]{\fQ}_\eta-\partial\fQ\right)^2\right>\nonumber\\
	=&\,\left<\left\{\dfrac{\lambda}{2}\left[{\fQ}_\eta\left(\theta+\frac{\pi}{2}\right)-{\fQ}_\eta\left(\theta-\frac{\pi}{2}\right)\right]-\partial\fQ\right\}^2\right>\nonumber\\
	=&\,\left<\left\{\dfrac{(1-\eta)\,\lambda}{2}\left[{\fQ}\left(\theta+\frac{\pi}{2}\right)-{\fQ}\left(\theta-\frac{\pi}{2}\right)\right]-\partial\fQ\right\}^2\right>\nonumber\\
	=&\,\left<\left[(1-\eta)\,\lambda\,\partial\fQ-\partial\fQ\right]^2\right>\nonumber\\
	=&\,\left[1-(1-\eta)\,\lambda\right]^2\left<|\partial\fQ|^2\right>\,.
\end{align}

The quadratic dependence in $\lambda$ for the SPS MSEs in \eqref{eq:MSE_SPS} permits the acquisition of explicit $N^{\mathrm{NSPS}}_*$ closed forms for \mbox{$g=0$}. When the total error rate of the noise channel is $\eta$, these \emph{exact} formulas are
\begin{align}
	N^{\mathrm{NSPS}}_{*}=&\,\dfrac{(d^2-1)h(d,\eta)}{2d^2\eta(1-\eta)}\quad\!\qquad(\text{$\partial \fQ$ estimation})\,,\nonumber\\
	N^{\mathrm{NSPS}}_{*}=&\,\dfrac{9(d^2-1)h(d,\eta)}{16d^2\eta(1-\eta)}\,\,\qquad(\text{$\partial\partial \fQ$ estimation})\,,\nonumber\\
	N^{\mathrm{NSPS}}_{*}=&\,\dfrac{(d^2-1)^2h(d,\eta)}{d^4\eta(1-\eta)}\,\,\qquad(\text{$\partial\partial' \fQ$ estimation})\,,\nonumber\\
	h(d,\eta)=&\,d+4\eta+\eta^2(d-2)\nonumber\\
	&\,+\big\{4\eta(2-\eta)^2+4d\eta\left[2+\eta(1-\eta)(3-\eta)\right]\nonumber\\
	&\,\quad\,\,+d^2\left[1+\eta\left(8-6\eta+\eta^3\right)\right]\big\}^{1/2}\,.
	\label{eq:SPS_N_star_full}
\end{align}
Clearly, if $\eta\rightarrow0$, $h(d,\eta)\rightarrow2d$ and we once again arrive at~\eqref{eq:SPS_N_star}.

\section{Biases in NSPS, NFD and HFD estimators}
\label{app:biases}

It is clear that using numerical estimators such as those of NSPS and NFD, which are optimized for noiseless quantum circuits, introduce noise biases, which are permanent systematic errors when used to estimate gradient and Hessian components for noisy circuit functions. 

To quantify these noise biases for NSPS and HSPS, we take the constant-$g$ and two-design (with the TDS condition) approximations and investigate things in the regime of large $N_\mathrm{T}$, where $\lambda_{\mathrm{opt}}\cong1$ for all NSPS schemes. Hence, while the finite-copy errors all go as $1/N_\mathrm{T}$, the approximation errors respectively approach $\left<|\partial\fQ|^2\right>\eta^2$, $\left<|\partial\partial\fQ|^2\right>\eta^2$ and $\left<|\partial\partial'\fQ|^2\right>\eta^2$, which are the noise biases. On the other hand, from the results of \eqref{eq:SPS_lbd_opt_eta}, we have the asymptotic answer $\lambda_{\mathrm{opt},\eta}\cong1/(1-\eta)>1$ for all HSPS schemes. It is then trivial to see that the approximation error approaches zero in the large-$N_\mathrm{T}$ limit. This reasoning works well so long as $g$ is approximately constant.

A similar argument may be invoked to study the noise biases of NFD estimators defined by the optimal parameters $\epsilon_{\mathrm{opt}}$ meant for noiseless circuits. In (C8) of \cite{Teo:2023optimized}, these parameters, in the $N_\mathrm{T}\gg d$ limit are found to be
\begin{align}
	\epsilon_\mathrm{opt}(\partial f_\mathrm{Q})\cong&\,\left[\dfrac{1152d}{\MEAN{(\partial f_{\mathrm{Q}})^2}{}N_\mathrm{T}(d+1)}\right]^{1/6}\,,\nonumber\\
	\epsilon_\mathrm{opt}(\partial\partial f_\mathrm{Q})\cong&\,\left[\dfrac{2592d}{\MEAN{(\partial\partial f_{\mathrm{Q}})^2}{}N_\mathrm{T}(d+1)}\right]^{1/8}\,,\nonumber\\
	\epsilon_\mathrm{opt}(\partial\partial' f_\mathrm{Q})\cong&\,\left[\dfrac{2304d}{\MEAN{(\partial\partial' f_{\mathrm{Q}})^2}{}N_\mathrm{T}(d+1)}\right]^{1/8}\,.
\end{align}
For such an astronomical $N_\mathrm{T}$, we return to the textbook optimality requirement that $\epsilon_{\mathrm{opt}}\rightarrow0$, so that $\sinc(\epsilon_{\mathrm{opt}}/2)\rightarrow1$. Notice that $\epsilon_\mathrm{opt}$ tends to zero much slower than $1/N_\mathrm{T}$, so that the finite-copy errors of \eqref{eq:MSE_FD} (evaluated with $\epsilon_{\mathrm{opt}}$ for NFD) still approach zero in a well-defined manner as $N_\mathrm{T}\rightarrow\infty$. Thus, the noise biases of NFD estimators are precisely those of NSPS estimators.

For the HFD estimators, it turns out that the approximation error can never be completely eliminated even in the limit of large $N_\mathrm{T}$. The straightforward reason is that when $\eta<1$, unlike $\lambda_{\mathrm{opt},\eta}$, which is to be greater than 1 for noise biases to vanish, the sinc functions entering the approximation errors of the FD estimators are all never greater than 1. Therefore, the best these HFD estimators can do is to achieve noise biases equal to those of the NFD estimators in the asymptotic limit.

\end{document}